\documentclass[12pt]{article}
\usepackage{
amsfonts,amssymb,epsfig,amsmath}
\usepackage{color}
\usepackage{hyperref,hep}
\usepackage{cancel}
\usepackage{mathrsfs}
\usepackage[letterpaper, text={6in,9in},centering]{geometry}

\renewcommand{\text}[1]{#1}

\newcommand{\be}{\begin{equation}}
\newcommand{\ee}{\end{equation}}
\newcommand{\ben}{\begin{displaymath}}
\newcommand{\een}{\end{displaymath}}
\newcommand{\bea}{\begin{eqnarray}}
\newcommand{\eea}{\end{eqnarray}}
\newcommand{\bean}{\begin{eqnarray*}}
\newcommand{\eean}{\end{eqnarray*}}
\newcommand{\nn}{\nonumber \\}
\newcommand{\ba}{\begin{array}}
\newcommand{\ea}{\end{array}}
\newcommand{\bi}{\begin{itemize}}
\newcommand{\ei}{\end{itemize}}

\def\Tr{{\rm Tr}}
\def\mS{\mathbb{S}}
\def\mH{\mathbb{H}}
\def\cS{{\cal S}}

\def\Xint#1{\mathchoice
   {\XXint\displaystyle\textstyle{#1}}%
   {\XXint\textstyle\scriptstyle{#1}}%
   {\XXint\scriptstyle\scriptscriptstyle{#1}}%
   {\XXint\scriptscriptstyle\scriptscriptstyle{#1}}%
   \!\int}
\def\XXint#1#2#3{{\setbox0=\hbox{$#1{#2#3}{\int}$}
     \vcenter{\hbox{$#2#3$}}\kern-.5\wd0}}

\def\dashint{\Xint-}

\newcommand{\es}[2] {\begin{equation} \label{#1} \begin{split} #2 \end{split} \end{equation}}

\begin{document}

\makeatletter
\renewcommand{\theequation}{\thesection.\arabic{equation}}
\@addtoreset{equation}{section}
\makeatother

\baselineskip 18pt

\begin{titlepage}

\vfill

\begin{flushright}
MIT-CTP/4579\\
\end{flushright}

\vfill

\begin{center}
   \baselineskip=24pt
   \begin{Large}\textbf{
        Super-R\'enyi Entropy \& Wilson Loops for ${\cal N} =4$~SYM and their Gravity Duals}
   \end{Large}
   \vskip 1.5cm
Michael Crossley$^{1)}$,  Ethan Dyer$^{1),2)}$ and Julian Sonner$^{1)}$ \\
   \vskip .6cm
 \begin{small}
     {\it 1) Center for Theoretical Physics, Massachusetts Institute of Technology\\ Cambridge, MA 02139, U.S.A.}
        \end{small}\\*[.6cm]
        
         \vskip .6cm
 \begin{small}
     {\it 2) Stanford Institute for Theoretical Physics, Department of Physics\\ Stanford University, Stanford, CA 94305, U.S.A.}
        \end{small}\\*[.6cm]

   \end{center}

\vfill

\begin{center}
\textbf{Abstract}
\end{center}

\begin{quote}
We compute the supersymmetric R\'enyi entropies across a spherical entanglement surface in ${\cal N}=4$  $SU(N)$ SYM theory using localization on the four-dimensional ellipsoid.  We extract the leading result at large $N$ and $\lambda$ and match its universal part to a gravity calculation involving a hyperbolically sliced supersymmetric black hole solution of ${\cal N}=4^+$ $SU(2)\times U(1)$ gauged supergravity in five dimensions. We repeat the analysis in the presence of a Wilson loop insertion and find again a perfect match with the dual string theory. Understanding the Wilson loop operator requires knowledge of the full ten-dimensional IIB supergravity solution which we elaborate upon.
\end{quote}

\vfill

\end{titlepage}

\setcounter{equation}{0}

\tableofcontents

\section{Introduction}
The study of the entanglement entropy, $S(A)$, of a region $A$ with respect to its complement in quantum field theory is marred by the difficulties involved in concrete calculations of this quantity. The entanglement entropy is defined as the von-Neumann entropy 
\be
S(A) = - \Tr  \rho_A \ln \rho_A\,,\qquad \rho_A = \Tr_{{\cal H}_{\overline A}}\rho
\ee
of the reduced density matrix, associated with a decomposition of the Hilbert space into degrees of freedom within the region $A$ and its complement $\bar{A}$, that is, ${\cal H} = {\cal H}_A\otimes {\cal H}_{\overline A}$. A common technique for obtaining $S(A)$ proceeds via the so called replica trick, whereby one first computes the traces of powers of the reduced density matrix, or equivalently, the $n-$th R\'enyi entropy
\be
S_n(A) = \frac{1}{1-n}\ln\Tr \rho_A^n\,.
\ee
One then obtains the entanglement entropy as the limit $S(A) = \lim_{n\rightarrow 1} S_n(A)$, after analytically continuing the R\'enyi entropy to non-integer $n$. The R\'enyi entropy at integer values of $n$ is obtained in terms of the partition function of the original theory on the $n$ replicated space, obtained by joining together $n$ copies of the original space along branch cuts on $\partial A$. We then have the R\'enyi entropy
\be\label{eq.singularPartitionFunction}
S_n(A) = \frac{1}{1-n}\ln\left[\frac{Z_n}{(Z_1)^n} \right]\,.
\ee
There are few examples of interacting quantum field theories where $Z_n$ can be calculated explicitly. For conformally invariant quantum field theories, such as the ${\cal N}=4$ $SU(N)$ SYM theory in $d=4$, the subject of this paper, Casini, Huerta and Myers \cite{Casini:2011kv} have shown that the computation of the R\'enyi entropies for the case of a ball-shaped region $A$ of radius $R$ with boundary $\partial A = \mS^{d-2}$ is equivalent to the computation of the free energy on the geometry $\mS_n^1 \times \mH^{d-1}$, a product of the circle $\mS^1_n$ with period $2\pi n$ and hyperbolic space of dimension $d-1$. This space, in turn, can be conformally mapped to the branched sphere $\mS_n^d$, which, however, has conical singularities. We shall describe the relevant geometries in more detail in the next section. This circle of ideas is of particular appeal in the context of the AdS/CFT correspondence, on the one hand because holographic duality allows the calculation of the the free energy on $\mS_n^1\times \mH^{d-1}$ in terms of the thermodynamics of topological black holes \cite{Myers:2010xs,Myers:2010tj,Casini:2011kv,Hung:2011nu}, and on the other hand, because an increasing body of work suggests a connection between entanglement entropy and the geometry of space-time \cite{Swingle:2009bg,VanRaamsdonk:2010pw}, an idea for which AdS/CFT provides a fertile testing ground. It is thus of great interest to study the properties of entanglement of quantum field theories with holographic duals. This paper contains two main calculations of such quantities. Firstly, we compute the supersymmetric R\'enyi entropy and construct its gravity dual in terms of a $\frac{1}{4}$-BPS hyperbolic black hole solution of $SU(2)\times U(1)$ gauged supergravity in 5D. Secondly we also compute the contribution to the supersymmetric R\'enyi entropy due to the insertion of a Wilson loop, and again construct its gravity dual by studying a fundamental string  in the black hole geometry.

In \cite{Pestun:2007rz}, it has been demonstrated that the technique of supersymmetric localization allows the computation of exact partition functions and expectation values of BPS operators\footnote{The development of these techniques was in fact initiated by the exact computations of BPS Wilson-loop expectation values using matrix models \cite{Erickson:2000af,Drukker:2000rr}.}, including examples when the theories are placed on non-trivial background geometries \cite{Hama:2012bg}. Unfortunately one faces an immediate difficulty in applying localization to the computation of $S_n(A)$, namely that the replicated space breaks all of the supersymmetry. It is however possible to switch on compensating background fields, such as sources for conserved R-symmetry currents, in order to restore half of the original supersymmetry and therefore enable a localization computation. Using this idea, \cite{Nishioka:2013haa} proposed a supersymmetric generalization of R\'enyi entropy for three-dimensional SCFTs, which was successfully matched to a dual gravity calculation \cite{Huang:2014gca,Nishioka:2014mwa} using a BPS hyperbolic black hole of ${\cal N}=2$ gauged supergravity in four bulk dimensions \cite{Romans:1991nq}. It was also noted in \cite{Nishioka:2013haa}  that the partition function on the $n$-replicated three-sphere (also referred to as the `branched' sphere), is exactly equivalent to the partition function of the three-dimensional ellipsoid $E^3(\ell,\tilde\ell)$ with squashing parameter $\sqrt{\ell/\tilde\ell}$. We can summarize these relationships as follows:
\be
\mS^1_n \times \mH^2\quad\longleftrightarrow\quad\mS_n^3 \quad \longleftrightarrow\quad E^3(\ell,\tilde\ell)\,.
\ee
The branched three-sphere at the center corresponds to the naively replicated geometry (\ref{eq.singularPartitionFunction}) with conical singularities. On the one hand it can be related by a Weyl rescaling to hyperbolic space, suggesting that the dual geometry is a topological black hole. On the other hand, the localization result on the singular $\mS_n^3$ coincides with that on the smooth space $E_b^3$. Thus one ends up in a situation where the holographic dual of a theory on the ellipsoid is given by a bulk solution with hyperbolic spatial slices, namely a hyperbolic black hole. For direct gravity duals of three-dimensional gauge theories on other kinds of ellipsoids see \cite{Martelli:2011fw,Martelli:2011fu}.

In this paper we apply a similar construction to the ${\cal N}=4$ theory in four dimensions. We construct a supersymmetric version of the R\'enyi entropy in terms of the partition function of the theory on an ellipsoid \cite{Hama:2012bg}. The related geometries are the same, lifted up by one dimension, viz.
\be
\mS^1_n \times \mH^3\quad\longleftrightarrow\quad\mS_n^4 \quad \longleftrightarrow\quad E^4(\ell,\tilde\ell)\,.
\ee
This then allows us to compute a supersymmetric generalization of the R\'enyi entropy to all orders in $N$ and the 't Hooft coupling $\lambda$ and to match the result in the supergravity limit $N\rightarrow \infty$, $\lambda \rightarrow \infty$ with a corresponding calculation in the hyperbolic black hole background. Here and throughout the paper we denote the super-R\'enyi entropy by $\cS_n(A)$.

In addition to the super-R\'enyi entropy itself, we also obtain the result after inserting a Wilson line in the fundamental representation, thus adding a non-trivial excitation to the `vacuum'. Again, the calculation can be reduced to a supersymmetric matrix model, whose large-$N,\lambda$ behavior is captured by a semi-classical bulk calculation. We consider a fundamental string which ends on the supersymmetric Wilson loop configuration at the boundary. It is essential for this calculation that we understand the ten-dimensional lift of the gravity solution, as it is not enough to simply study the five-dimensional effective theory. This illustrates an interesting limitation of the use of lower-dimensional supergravities in precision holography: for certain observables the full ten-dimensional geometry is necessary in order to capture the dual field theory.

 In  four dimensions the universal part of the supersymmetric R\'enyi entropy is related\footnote{We thank Silviu Pufu for pointing this out to us.}  to the Weyl anomaly\cite{Capper:1974ic,Henningson:1998gx} on the four-dimensional ellipsoid, which can therefore easily be deduced from the localization result. This should be contrasted with the situation for the ordinary R\'enyi entropy \cite{Lewkowycz:2014jia}.  On the other hand this also means that one could determine the universal part of the super-R\'enyi entropy without a full localization calculation. However, no such shortcut seems to exist for the Wilson-loop and the full partition function is required in order to find its contribution to the supersymmetric R\'enyi entropy. 

This paper is structured as follows. In section 2 we calculate the supersymmetric R\'enyi entropy of ${\cal N}=4$ theory, as the partition function of the same theory on the four-dimensional ellipsoid. We then take the supergravity limit of the resulting partition function. In section 3 we compute the free energy of certain BPS black holes with hyperbolic horizons and match the result to that obtained in section 2. We conclude in section 4 with a discussion of our results. There are two appendices, appendix A contains the details of the BPS analysis of the $SU(2)$ black hole solution of ${\cal N}=4^+$ $SU(2)\times U(1)$ gauged supergravity, while appendix B contains details of the embedding of that solution into the ${\cal N}=2$ STU model.

Note added: as we were finishing this paper, we became aware of a preprint \cite{Huang:2014pda}, in which the authors also obtain the super-R\'enyi entropy from localization, matching the result with a supergravity dual in the ${\cal N}=2$ STU model. We explain the detailed relation to our dual in appendix \ref{sec.STU}.

\section{Computing $\cS_n$ in ${\cal N}=4$ SYM}
In this section we describe the computation of supersymmetric R\'enyi entropy in the $\mathcal{N}=4$ Super Yang-Mills theory. We first describe the field theory setup and then explain the geometries this theory is studied on, namely the branched sphere and the ellipsoid. We go on to expand the result for the localized partition function and Wilson loop on the ellipsoid at large $N, \lambda$ for gauge group $SU(N)$.
\subsection{Field Theory Setup}\label{sec.FieldTheorySetup}
As mentioned above, the R\'enyi entropy of a circular region in a $d$ dimensional conformal field theory can be computed by calculating the Euclidean path integral for the theory on a branched $d$-sphere \cite{Casini:2011kv,Klebanov:2011uf}.
\es{bsphr}{
ds^{2}_{\mS^{d}_{n}}&=\left(d\theta^{2}+n^{2}\cos^{2}(\theta)d\tau^{2}+\sin^{2}(\theta)d\Omega^{2}_{d-2}\right)\,.
}

\noindent Here $\theta\in[0,\frac{\pi}{2}]$, $\tau\in[0,2\pi)$, and we have set the radius of the sphere to 1. This space is smooth for $n=1$, and singular otherwise. We will be interested in the case of four dimensions, $d=4$.

The $\mathcal{N}=4$ theory with gauge group $G$ consists of a vector field, $A_{\mu}$, six scalars, $\phi^{I}$, and four two component Weyl fermions, $\psi^{a}$. Each of these fields are matrices with the indices suppressed, and take values in the Lie algebra of $G$. The theory has an $SU(4)$ R-symmetry, under which the scalars, $\phi^{I}$, transform in the $\mathbf{6}$ (the vector of $SO(6)$) and the fermions transform in the fundamental. The action is given by:
\es{SYM}{
S_{\mathcal{N}=4}=-\frac{1}{g^{2}_{\rm YM}}\int d^{4}x\Tr\left(\frac{1}{2}F_{\mu\nu}F^{\mu\nu}+D_{\mu}\phi_{I}D^{\mu}\phi^{I}+\frac{1}{2}\sum_{I,J}[\phi^{I},\phi^{J}]^{2}+\ldots\right)\,.
}
We will be interested in the case of $G=SU(N)$, for which the theory has a bonus, $U(1)_{B}$, symmetry at large $N$ and 't Hooft coupling \cite{Intriligator:1998ig,Intriligator:1999ff}.

This theory has 16 real supercharges, excluding the superconformal charges, and it is possible to compute the Euclidean partition function exactly on $\mS^{4}$ \cite{Pestun:2007rz}. We are not only interested in the partition function for the round sphere, but for the singular space, (\ref{bsphr}).

\subsubsection{Branched is Squashed}\label{sec.branchedIsSquashed}
To understand how to evaluate the partition function on the branched four sphere, let's first review a similar computation in three dimensions. In \cite{Nishioka:2013haa,Huang:2014gca}, the authors were able to compute the Euclidean partition function for $\mathcal{N}\geq2$ SCFTs on a smoothed version of the branched three sphere,
\es{ssphr3}{
ds^{2}_{\mS^{3}_{n,\epsilon}}&=\left(f_{\epsilon}(\theta)d\theta^{2}+n^{2}\cos^{2}(\theta)d\tau^{2}+\sin^{2}(\theta)d\phi^{2}\right)\,.
}

\noindent Here, $f_{\epsilon}$ is a function which gives a smooth space for any $\epsilon>0$ and reduces to the branched sphere for $\epsilon=0$. The localization result is independent of $\epsilon$, and thus can be thought of as computing the partition function on the singular space. This is an example of the wider phenomenon of deformation independence of localized partition functions, as studied for example in \cite{Alday:2013lba,Closset:2013vra, Gerchkovitz:2014gta}.

Not only is the result independent of the smoothing parameter, $\epsilon$, but it is also equal to the partition function on a particular squashed three sphere \cite{Hama:2011ea}, given by the equation
 \es{sqsphr3eq}{
\frac{x_{1}^{2}+x_{2}^{2}}{\ell^2}+\frac{x_{3}^{2}+x_{4}^{2}}{\tilde{\ell}^2}&=1\,
}

\noindent with metric,
\es{sqsphr3}{
ds^{2}_{E^{3}(\ell,\tilde\ell)}&=\left(f(\theta)d\theta^{2}+\ell^{2}\cos^{2}(\theta)d\tau^{2}+\tilde{\ell}^{2}\sin^{2}(\theta)d\phi^{2}\right)\,,
}
where $f(\theta)=\ell^{2}\sin^{2}(\theta)+\tilde{\ell}^{2}\cos^{2}(\theta)$. To make contact with the branched sphere we can take $\ell=n$ and $\tilde{\ell}=1$.

In summary, the partition function on the branched three sphere is equal to that on the squashed three sphere, (\ref{sqsphr3eq}).

In four dimensions, the geometry is quite similar. The branched four-sphere can be thought of as a singular fibration of the branched three-sphere over an interval. For this reason we identify the R\'enyi entropy with the partition function on the squashed four sphere, given by\footnote{During the completion of this paper \cite{Huang:2014pda} proved that the squashed four-sphere partition function is equal to a smoothed four-sphere partition function, and that the result is independent of the smoothing parameter, validating the identification.}:
 \es{sqsphr4eq}{
\frac{x_{1}^{2}+x_{2}^{2}}{\ell^2}+\frac{x_{3}^{2}+x_{4}^{2}}{\tilde{\ell}^2}+\frac{x_{5}^{2}}{r^2}&=1\,,
}  
where we take, $\ell=n$, $\tilde{\ell}=1$, and $r=1$ to recover the result on the branched sphere.
Fortunately the localization of the partition function on this space was carried out for arbitrary $\mathcal{N}=2$ gauge theories in \cite{Hama:2012bg}.
\subsubsection{The Squashed Theory}\label{sec.squashedTheory}
In order to put the $\mathcal{N}=4$ theory on the ellipsoid, (\ref{sqsphr4eq}), it is necessary to turn on various background fields. This breaks the $SU(4)$ R-symmetry to a subgroup. To see how this works, it is easiest to split the fields into an $\mathcal{N}=2$ vector, and a hyper multiplet. 

The vector multiplet consists of a vector, a complex scalar, and two Weyl fermions, as well as their formal conjugates: $(A_{\mu},\phi,\tilde{\phi},\lambda_{A},\tilde{\lambda}_{A})$. Here, $A=1,2$ and $\lambda_{A}$ is in the fundamental of an $SU(2)_{V}$ subgroup of the R-symmetry group. 
The hypermultiplet contains the remaining four scalars, and two Weyl fermions. They can be packaged as $(q_{IA},\psi_{I},\tilde{\psi}_{I})$. The index $I=1,2$ and indicates an $SU(2)_{H}$ subgroup of the R-symmetry group.
\begin{table}[t!]
\centering 
\begin{tabular}{c || c |c |c|} 
& $SU(2)_V$ & $SU(2)_H$ & $U(1)_R$ \\ [0.5ex] 
\hline\hline
$A_\mu$ & 0 & 0 & 0 \\ 
$\phi$& 0 & 0 & 2 \\
$\lambda_A$ & $\frac{1}{2}$ & 0 & 1 \\
\hline
$q_{IA}$ & $\frac{1}{2}$ & $\frac{1}{2}$ & 0 \\
$\psi_I$ & 0 & $\frac{1}{2}$ & -1 \\ [1ex] 
\hline 
\end{tabular}
\caption{R-charge assignments of the various fields on the ellipsoid. The formal conjugates have opposite charges under $U(1)_R$.} \label{tab.Rcharges}
\end{table}

In order to put this theory on the squashed four-sphere, one must introduce couplings to several background fields, $(M,(V_{\mu})^{A}_{\ B},T^{\mu\nu},\tilde{T}^{\mu\nu})$. These background fields are coupled through the interaction terms:

\es{bgint}{
\mathcal{L}_{\rm Vbg}&=\Tr\left(M\tilde{\phi}\phi+16 F_{\mu\nu}\tilde{\phi}T^{\mu\nu}+64\tilde{\phi}^{2}T_{\mu\nu}T^{\mu\nu}+\ldots\right) + \textrm{conjugate}\\
\mathcal{L}_{\rm Hbg}&=\Tr\left(\frac{1}{8}(1+M)q_{IA}q^{IA}-q_{IB}q^{I}_{C}(V_{\mu})^{B}_{\ A}(V^{\mu})^{CA}+\frac{i}{2}q_{IA}D^{\mu}q^{I}_{B}(V_{\mu}^{AB}+V_{\mu}^{BA})\ldots\right)\,.
}
The ellipses indicate fermionic couplings. The round sphere corresponds to $M=-1/3$, with vanishing $T,\tilde{T}, V$. These couplings preserve an  $SU(2)_{V}\times SU(2)_{H}\times U(1)_{R}$ R-symmetry, where the complex fields $(\phi,\lambda_{A},\psi_{I})$ have $U(1)_{R}$ charges $(2,1,-1)$, the formal conjugates have the opposite charge, and the remaining fields are neutral.

On the squashed sphere, however, $T$, $\tilde{T}$, and $V$ are non-vanishing, and thus only the $SU(2)_{H}$ remains,
\be\label{eq.unbrokenRsymmetry}
SU(4)_R\longrightarrow SU(2)_H  \xrightarrow{\textrm enhanced} SU(2)_H \times U(1)_B\,.
\ee
 This unbroken $SU(2)_{H}$, enhanced by the bonus $U(1)_{B}$,  will be useful in identifying the bulk dual in section \ref{sec.BulkDual}. The charge assignments are summarized in table \ref{tab.Rcharges}.

The authors of \cite{Hama:2012bg} were able to use such background couplings for arbitrary $\mathcal{N}=2$ gauge theories and compute the localized partition function\footnote{For a pedagogical review of these techniques, albeit with a focus on three dimensions, see \cite{Marino:2011nm}.}. For the particular case of $\mathcal{N}=4$ SYM, the partition function is given by:
\es{partf1}{
Z&=\int dae^{-\frac{8\pi^{2}}{g_{YM}^{2}}\Tr(a^{2})}|Z_{\rm inst}|^{2}\prod_{\alpha\in\Delta^{+}}\frac{\Upsilon(ia\cdot\alpha)\Upsilon(-ia\cdot\alpha)}{\Upsilon(ia\cdot\alpha+\frac{Q}{2})^{2}}\,.
}

Here, $\alpha$ runs over the positive roots of the gauge group, the matrix $a=a_{i}h_{i}$ parameterizes the Cartan of $G$, $|Z_{\rm inst}|^{2}$ is the Nekrasov partition function, \cite{Nekrasov:2002qd}, $Q=\sqrt{n}+1/\sqrt{n}$, and we have introduced the function

\es{Upsilon}{
\Upsilon(x)&=\prod_{p,q\geq0}\left(p\sqrt{n}+\frac{q}{\sqrt{n}}+Q-x\right)\left(p\sqrt{n}+\frac{q}{\sqrt{n}}+x\right)\,.
}

In fact, The ratio in (\ref{partf1}) is divergent in a way that cannot be removed be the addition of a local counter term. In the free energy, $F=-T \log Z$, this divergence shows up as a logarithmic divergence,
\be
F\supset a_n\log(R/\delta)\,,\qquad a_{n}=\frac{Q^{2}}{4} a_{1}\,,
\ee
where $a_n$ is a universal constant related to the Weyl anomaly and the explicit expression for $a_n$ applies to the ellipsoid.
We will be interested in the case $G=SU(N)$ for which (\ref{partf}) becomes:

\es{partf}{
Z&=\int \prod_{i=1}^{N-1}da_{i=1}e^{-\frac{8\pi^{2}N}{\lambda}\sum_{i=1}^Na_{i}^{2}}|Z_{\rm inst}|^{2}\prod_{i<j}\frac{\Upsilon(i a_{ij})\Upsilon(i a_{ji})}{\Upsilon(i a_{ij}+\frac{Q}{2})^{2}}\,,
}
where $a_{ij} = a_i - a_j$ and we have introduced $a_N = -\sum_{i=1}^{N-1}a_i$.
\subsection{The Partition Function at large $N$}
We will follow the procedure outlined in \cite{Russo:2012ay,Buchel:2013id,Russo:2013sba} to obtain the large $N$, $\lambda$ limit of the localization result of $SU(N)$ $\cal{N}=$4 SYM on the ellipsoid obtained in section \ref{sec.FieldTheorySetup}. 
In the large $N$ limit, the integral (\ref{partf}) can be evaluated exactly in the saddle point approximation, and the contributions of $Z_{\rm inst}$ are exponentially suppressed, as in the ${\cal N}=2^*$ theory \cite{Russo:2012ay,Buchel:2013id,Russo:2013sba,Billo:2014bja}. The saddle-point equations\footnote{There are subleading in $N$ contributions that arise from the $SU(N)$ constraint, which we suppress here.} are
\be
\sum_{j\neq i}\left( K\left(a_{ij}\right)- K\left(a_{ji}\right) - 2K\left(a_{ij}-i\tfrac{Q}{2}\right)\right)=\frac{16\pi^2N}{\lambda} a_i,
\ee
where 
\begin{align}
K(x)&=i\frac{\Upsilon'(ix)}{\Upsilon(ix)}\nn
&=i\sum_{p,q=0}^\infty \frac{Q-2i x}{\left( \sqrt{n}p + \frac{q}{\sqrt{n}} + Q - ix  \right) \left(   \sqrt{n}p + \frac{q}{\sqrt{n}} + i x   \right)}\,.
\end{align}
In the large $N$ continuum limit, the saddle-point equation becomes an integral equation for the eigenvalue density $\rho(y)$:
\be
\dashint_{-\mu}^{\mu}\rho(y)\left( K\left(x-y\right) - K\left(y-x\right)  - 2K\left(x-y-i\tfrac{Q}{2}\right)\right)dy=\frac{16\pi^2}{\lambda} x,
\ee
where $\dashint_{-\mu}^{\mu}$ denotes the principal-value integral. As in \cite{Russo:2012ay,Buchel:2013id,Russo:2013sba}, we assume that the symmetry breaking scale is much larger than the deformation, so that $\mu\gg Q$. We will see below that this holds for large $\lambda$. In this case, the integral will be dominated by the leading behavior in $\frac{1}{x}$. Expanding the integrand about $x=\infty$ yields for the integral equation
\be
\dashint_{-\mu}^{\mu}\rho(y)\frac{Q^2}{4\left(x-y\right)}dy=\frac{8\pi^2}{\lambda}x.
\ee
The solution is given by the Wigner semi-circle distribution,
\be
\rho(x)=\frac{2}{\pi \mu^2}\sqrt{\mu^2-x^2},
\ee
where
\be
\mu=\frac{\sqrt{\lambda}Q}{4\pi}.
\ee
We see that the assumption that $\mu\gg Q$ is justified for large $\lambda$. With this eigenvalue distribution, we can compute the value of the partition function, yielding
\be\label{eq.FreeEnergyFieldTheory}
F = - N^2\frac{Q^2}{4}\log\left(R/\delta \right) -\frac{N^2}{8}Q^2\ln\lambda+\ldots,
\ee
where we have restored the divergence proportional to the universal constant $a_n$ and the neglected terms are subleading in $\lambda$. Though the second term in (\ref{eq.FreeEnergyFieldTheory}) is scheme-dependent, we can compare to a gravity calculation, by choosing on both sides the particular scheme advocated in \cite{Russo:2012ay,Buchel:2013id,Russo:2013sba}. This allows us to precisely match both the universal part and the leading $\lambda$ dependence below.
Finally, the  large $N,\lambda$ scaling of the free energy is
\be
F_n=\frac{\left(n+1\right)^2}{4n}F_1.
\ee
\subsection{Wilson loop}
The localization procedure goes through also in the presence of a BPS Wilson loop operator.
The localized $\varphi$ Wilson loop, in the nomenclature of  \cite{Hama:2012bg}, is given by an insertion of 
\be
W_R = \mbox{Tr}_R \exp\left(-2\pi \sqrt{n} a\right)
\ee
in the localized partition function for a given representation $R$. For $\mathcal{N}=4$ SYM on the ellipsoid at large $N$, $\lambda$, we then have, for a Wilson loop in the fundamental representation,
\be
\langle W\rangle=\dashint_{-\mu}^\mu dx\rho(x)e^{-2\pi \sqrt{n} x}.
\ee
The integral evaluates to
\be\label{eq.WilsonLoopMatrix}
\langle W\rangle=\sqrt{\frac{2}{\pi}}\left(\lambda n\frac{Q^2}{4}\right)^{-\frac{3}{4}}e^{\sqrt{\lambda n}\frac{Q}{2}}+\ldots,
\ee
yielding
\be
\ln W_n = \frac{n+1}{2}\sqrt{\lambda}
\ee
and the corresponding scaling relation,
\be
\ln W_n = \frac{(n+1)}{2}\ln W_1.
\ee
\subsection{Supersymmetric R\'enyi entropies}\label{sec.SUSYRenyi}
Substituting into the definition of the supersymmetric R\'enyi entropy given in section 1, we find, in the absence of the Wilson loop, from the result for $F_n$ given in subsection 2.2.1,
\be
S_n=-\frac{3n+1}{4n}F_1.
\ee
In the presence of the Wilson loop, the additional contribution to the supersymmetric R\'enyi entropy, from the result for $\ln W_n$ given in subsection 2.2.2, is
\be
S^{(W)}_n=\frac{1}{2}\ln W_1,
\ee
independently of $n$. All the results obtained in this section for the $n$-scaling of quantities are the same as those of the analogous quantities in three dimensions found in \cite{Huang:2014gca, Nishioka:2013haa}.
We can compare this result with \cite{Lewkowycz:2013laa}. Using the formula obtained there, our $\log W_1$ gives for the entanglement entropy of the Wilson loop
\begin{align}
S_{EE}&=\left(1-\frac{4}{3}\lambda\partial_\lambda\right)\log W_1\nn
&=\frac{1}{3}\sqrt{\lambda}
\end{align}
in agreement with the expression obtained in \cite{Lewkowycz:2013laa}, and previously in \cite{Jensen:2013ora}, using results of \cite{Jensen:2013lxa} (see also \cite{Sonner:2013mba}). If we calculate the $n\rightarrow1$ limit of our super-R\'enyi entropy, however, we find
\be
S^{(W)}_{1}=\frac{1}{2}\sqrt{\lambda}.
\ee
These results are not guaranteed to coincide, as the limits involve the first derivative away from $n=1$ of the Wilson loop action. In the entanglement entropy calculation, this derivative is taken with the charge fixed to be zero, where as in our supersymmetric background, a non-trivial charge is switched on.
\section{The gravity dual of $\cS_n$}\label{sec.BulkDual}
In section \ref{sec.squashedTheory} we identified the unbroken global symmetries of the original $SU(4) \cong SO(6)$ R-symmetry of ${\cal N}=4$ SYM theory on the ellipsoid to be $SU(2)_H$, which gets enhanced by the bonus symmetry to $SU(2)_H\times U(1)_{B}$. We now want to find a bulk dual which holographically encodes this pattern. It is well-known that boundary global symmetries are represented by local symmetries in the  bulk. We will now describe how this works in detail in our case.
\subsection{$ SU(2) \times U(1)$ gauged Supergravity}
Given the global symmetries on the boundary, (\ref{eq.unbrokenRsymmetry}), the natural candidate dual is the ${\cal N} = 4$ supersymmetric gauged supergravity theory with $SU(2)\times U(1)$ local symmetry in five dimensions, which was derived by Romans in \cite{Romans:1985ps}, and denoted ${\cal N}=4^+$. This theory can be viewed as a sub truncation of the $SO(6)$-gauged ${\cal N}=8$ supergravity, which is known to be dual to the ${\cal N}=4$ SYM theory with unbroken R-symmetry on the boundary. Thus the way the ${\cal N}=4^+$ bulk theory sits in the $SO(6)$ theory is exactly the bulk dual of the global symmetry pattern resulting from putting ${\cal N}=4$ SYM on the ellipsoid in the supergravity limit. The theory can be lifted to type IIB theory in ten dimensions, as shown in \cite{Lu:1999bw}, relying on previous results in \cite{Cvetic:1999xp}. We shall use the conventions of these references. 

The ${\cal N}=4^+$ supersymmetry is carried by four $USp(4)$ symplectic Majorana supercharges with a total of 16 real degrees of freedom.
The bosonic sector of the five-dimensional ${\cal N}=4^+$ theory contains a metric $g$, a scalar field $X$, a $U(1)$ gauge field $B_\mu$, an $SU(2)$ gauge field $A_\mu^I$, and two 2-form tensor fields $T_{\mu\nu}^a$, where $a$ is a doublet index under $U(1)$. We note that this field content exactly mirrors the background multiplet needed to put ${\cal N}=4$ SYM on the ellipsoid \cite{Hama:2012bg}. Here we show that the theory contains black hole solutions, charged under the $SU(2)$, which we will construct now. These solutions have $T^a_{\mu\nu}=0$ and $B_\mu=0$, so it suffices to consider only the action for the remaining fields. This takes the simple form
\be\label{eq.N=4plusLag}
S =\frac{1}{16 \pi G_N^{(5)}} \int d^5 x \sqrt{-g} \left(R  + g^2 \,{\cal V} - 3 X^{-2} \partial_\mu X \partial^\mu X  - \frac{1}{4}X^{-2} F_{\mu\nu}^I F^{I\,\mu\nu}  \right) +\cdots\,,
\ee
where the omitted terms include a Chern-Simons term for the gauge field and the fermions. These do no not contribute to any of the solutions in this paper. Until further notice we work in units where $16 \pi G_N^{(5)}=1$. Then the scalar potential reads
\be
{\cal V} = 4(X^2 + 2 X^{-1})\,.
\ee
This theory has a particular black-hole solution, which asymptotes to hyperbolically sliced $AdS_5$. The metric is
\be\label{eq.STUBlackHole}
ds_5^2 = -\frac{f(r)}{H(r)^{4/3}}dt^2 +  H(r)^{2/3} \left(\frac{dr^2}{f(r)} + r^2 ds^2(\mH^3)\right)\,,
\ee
where the functions $f$ and $H$, as well as the scalar field $X$ are given by
\be
f(r) = -1 -\frac{m}{r^2}+ g^2 r^2 H^2\,,\qquad X = H^{-1/3}\,,\qquad H(r) = 1 + \frac{q}{r^2}\,.
\ee
For later use, let us write down the line element of $\mathbb{H}^3$ we work with:
\be\label{eq.hyperbolicLineElement}
ds^2 (\mathbb{H}^3) = du^2 + \sinh^2 u \,d\Omega_2^2\,.
\ee
We can choose a gauge such that $A^I\Gamma_I$ lies entirely in the direction of the Cartan element of $SU(2)$,
\bea
A^{I=3} &=&\left[i\sqrt{2(1-\tfrac{m}{q})} (1-H^{-1})   + \mu \right]dt  \nn
&=& \frac{i\sqrt{2q(q-m)}}{r^2 + q}dt + \mu dt\,,\qquad A^I = 0 \,,\qquad (I=1,2) \,,
\eea
where the chemical potential $\mu$ is determined by requiring the gauge field to vanish at the horizon.
\subsection{Uplift to Ten Dimensions}
In \cite{Lu:1999bw} it is shown how to obtain the full ${\cal N}=4^+$ from a non-abelian sphere reduction of type IIB theory in ten dimensions. Here we only need the special case which lifts the fields active in the solution (\ref{eq.N=4plusLag}) to type IIB, which is considerably simpler. The ten dimensional metric is
\be\label{eq.SU2Uplift}
ds^2_{10} = \sqrt{\Delta}ds^2_5 +g^{-2} X\sqrt{\Delta} d\xi^2 + g^{-2}\Delta^{-1/2}X^2 s^2 d\tau^2 +\tfrac{1}{4}g^{-2}\Delta^{-1/2} X^{-1}c^2 \sum_{I=1}^3 \left( \sigma^I - \sqrt{2} g A^{I}\right)^2\,,
\ee
where $\tau$ and $\xi$ are angles on the $S^5$, $s = \sin\xi$ and $c = \cos\xi$ and $\sigma^I$ are the standard left-invariant one-forms on $SU(2)$ parametrizing the remaining three angles on $S^5$. They can be expressed in terms of the Euler angles as
\begin{align}
\sigma_1+i \sigma_2&=e^{-i\psi}\left(d\theta+i\sin\theta d\phi\right),\nn
\sigma_3&=d\psi+\cos\theta d\phi.
\end{align}
We then have
\be
\Delta = X^{-2}s^2 + X c^2\,.
\ee
The solution has vanishing axi-dilaton, and we do not need the functional form of the remaining fields. They are given in detail in \cite{Lu:1999bw}. 

Note that in the special case when the gauge field is in the Cartan direction, the uplift given becomes identical to that of the $U(1)^3$ abelian gauged supergravity in five dimensions \cite{Cvetic:1999xp}, used in \cite{Huang:2014pda}.

We give the details of this embedding in appendix \ref{sec.STUembedding}.
\subsection{BPS Analysis}
In order to determine the supersymmetry properties of this solution, we must study the fermion variations. Let $\varepsilon$ be a $USp(4)$ symplectic Majorana spinor. The fermion variations are given in detail in appendix \ref{app.SUSY}, consisting of the spin-1/2 gaugino and the spin-3/2 gravitino variation. Here we write the gaugino variation, (\ref{eq.FermiVariations}) in the schematic form
\be
{\cal M}(q,m)_a{}^b \varepsilon_b = 0\,.
\ee
For this linear equation to admit non-trivial solutions, we must demand that the determinant of the operator ${\cal M}$ vanish.

%
%


After some algebra one can show that for $q\neq 0$ this equation implies the condition
\be\label{eq.Projector}
0 = P_a{}^b\varepsilon_b = \frac{1}{2} \left(\delta_a^b + \frac{1}{\sqrt{f}} \left[ \gamma_0\otimes (\Gamma_3)_a{}^b + i g r H \gamma_1 \otimes (\Gamma_{45})_a{}^b \right]\right)\varepsilon_b\,,
\ee
where $\gamma_{0,1}$ are tangent-space gamma matrices. Computing its determinant, we find
\be
({\rm det} \,P)^{1/8}  =\frac{m}{r^2 f(r)}\,,
\ee
so that nontrivial solutions must have $m=0$.

The condition (\ref{eq.Projector}) projects out half of the components of the spinor $\varepsilon$. We show in appendix \ref{app.SUSY} that the gravitino variation imposes a further projection condition
\be
\varepsilon = i \Gamma_{345}\varepsilon\,,
\ee 
so that the black hole is a $\tfrac{1}{4}$-BPS solution, preserving four real supercharges. The same amount is preserved by the ${\cal N}=4$ SYM theory on the ellipsoid \cite{Hama:2012bg}. We give a detailed treatment of the supersymmetry of this solution including the full set of preserved Killing spinors in appendix \ref{app.SUSY}.

We next record some useful facts about the BPS solution. The horizon position is determined by the the equation $f(r_h)=0$, which gives rise to the condition
\be\label{eq.qtorh}
q = - \frac{r_h (gr_h \pm 1) }{g}.
\ee
The Hawking temperature is found with the usual Euclidean methods to be
\be\label{eq.2chargeTHawking}
T = \frac{g^2 }{2\pi r_h}\left( r_h^2 - q \right)\,.
\ee
We will later need the solution at some reference temperature $T_n =T_0/n\,, $
where $T_0 = \lim_{q\rightarrow 0}T$. Hence, by combining (\ref{eq.qtorh}) and (\ref{eq.2chargeTHawking})  we can find a relation between the horizon radius and the replica index $n$,
\be
r_h = \frac{n\pm1}{2gn}\,.
\ee
Interestingly this takes the same form as in four bulk dimensions \cite{Nishioka:2014mwa}, the significance of which will become apparent soon. Since $g r_h$ should go to unity, when $n=1$, we choose the positive branch. The same reasoning reveals that we should choose the minus sign in (\ref{eq.qtorh}) in order to get $gr_h=1$ when $n=1$. Finally we can express the charge in terms of the replica index as
\be
q = \frac{n^2 -1 }{4g^2 n^2}\,.
\ee

\subsection{The Euclidean Action}\label{sec.ActionMatch}
Defining the  Gibbs free energy $F(\beta,\mu) =  -T \ln {\cal Z}(\beta, \mu)$ via the usual grand canonical partition function ${\cal Z}(\beta,\mu) = e^{-I_E}$, we have
\be
F(\beta,\mu) = I_E(\beta,\mu)\,,
\ee
so we need to compute the Euclidean partition function of the dual bulk gravity.
In order to compute the free energy holographically, we need to evaluate the on-shell value of the Euclidean action. The analytic continuation is achieved by setting $t = -i\tau$, so that $-iS = -I_E$. 
To the action (\ref{eq.N=4plusLag}) we first add the Gibbons-Hawking boundary term, in order to have a well-defined variational principle. As usual the action contains UV divergences, which are dealt with by introducing UV counter terms. One obtains a finite expression
\be
I_E = \lim_{\Lambda\rightarrow\infty} \left[ \int_{r_h}^\Lambda d^5x \sqrt{g_E}{\mathscr L}_E + I_{\rm GH}(\Lambda) + I_{\rm ct}(\Lambda)\right]\,,
\ee
with $I_{\rm GH}$ the Gibbons-Hawking term, and $\Lambda$ is a UV regulator. The terms $I_{\rm ct}$ being given by the usual expression for $5D$ bulk gravity \cite{Balasubramanian:1999re}
\be
I_{\rm ct} = 3 g \int d^4 x\sqrt{\gamma}\left[1 + \frac{1}{12g^2}R_{\rm b} \right]\,,
\ee
where $R_{\rm b}$ is the curvature scalar of the induced metric. We now compute the action on the solution itself. For this it is useful to observe that one may schematically write the bosonic part of the supergravity action (\ref{eq.N=4plusLag}) as
\be
S = \int_{r_h}^\Lambda \sqrt{g_E}\left(R + {\mathscr L}_m  \right)
\ee
so that the Einstein equation becomes
\be
R_{\mu\nu} - \frac{1}{2}R g_{\mu\nu} = \frac{1}{2}{\mathscr L}_m g_{\mu\nu} + \hat T_{\mu\nu}\,.
\ee
On the right hand side we split up the energy momentum tensor into the piece that comes from varying the $\sqrt{g_E}$ part of the action and the {\it explicit} variation of ${\mathscr L}_m$ with respect to the metric, denoted $\hat T_{\mu\nu}$ . It is easy to see that for a solution of the form (\ref{eq.STUBlackHole}) the $\mathbb{H}^3$ components of $\hat T$ vanish identically. We may thus use the $uu$ component of the Einstein equation to find
\be
{\mathscr L}_m = -R +2 g^{uu}R_{uu}\,.
\ee
Note that at an intermediate stage we have also used that the metric is diagonal. Now the full on-shell action reads simply
\be
S_{\rm on shell} = 2\int \sqrt{g_{E}} g^{uu}R_{uu} d^5x\,.
\ee
A brief computation of the relevant curvature term for the $\tfrac{1}{4}$ BPS black hole (\ref{eq.STUBlackHole}) reveals that we can write this integral as a bulk term plus a total derivative
\be
S_{\rm on shell} = -2\beta {\rm vol}  \left( \mathbb{H}^3 \right) \left[ \int_{r_h}^\Lambda 2r dr  + \int_{r_h}^\Lambda \partial_r \left(r^2 f(r) +\frac{1}{3} r^3 f(r)\frac{H'(r)}{H(r)} \right) dr \right]\,,
\ee
where $\beta = 2\pi n/g$.
The total derivative terms give no contribution from the lower end of the integral due to the fact that $f(r_h)=0$, so their contributions come entirely from the UV end of the integral. Together with the UV contribution from the first term these are cancelled by the UV counter terms. Thus the action becomes
\be\label{eq.EuclideanAction}
I_n = 4\pi ng^{-1} {\rm vol}  \left( \mathbb{H}^3 \right) r_h^2 = 4\pi {\rm vol} \left( \mathbb{H}^3 \right)  \frac{(n+1)^2}{4g^3n}\,.
\ee
As explained, for example in \cite{Hung:2011nu}, the volume of hyperbolic space is divergent, in accordance with the usual power law divergence for entanglement entropies. In odd dimensions, there is a subleading logarithmic term  \cite{Hung:2011nu}, and it is the coefficient of this term, which is universal. In our particular case (recall we use the line element (\ref{eq.hyperbolicLineElement}))
\be
{\rm vol}(\mathbb{H}^3)_{\rm univ} = - 2 \pi \ln \left(  2 R/\delta_{\rm SG} \right)\,.
\ee
Here $R$ is the radius of the spherical entangling surface\footnote{Strictly speaking, in our computation $R=L_{\rm AdS} = g^{-1}$ as can be seen from the metric (\ref{eq.STUBlackHole}), but also by tracing through the details of the standard conformal map in  \cite{Casini:2011kv}. However, as further shown in \cite{Casini:2011kv}, with a bit of care, one can incorporate an arbitrary size interval $R$ in the conformal map to $\mS_n^1 \times \mathbb{H}^3$. The universal scaling result  (\ref{eq.GravityUniversalScaling}) follows irrespective of whether $R=g^{-1}$ or is kept arbitrary.} and $\delta_{\rm SG}\ll R$ is a UV scale in the supergravity calculation, defined by cutting off the integral over hyperbolic space at some maximum $\cosh u_{\rm max} = R/\delta_{\rm SG}$.
Note finally that we can rewrite this in the familiar form
\be
I_n =- \frac{1}{4}Q^2 \frac{ \pi }{2 g^3 G_N^{(5)}} \ln \left(  2 R/\delta_{\rm SG} \right)\,,
\ee
where we  also have restored dimensionful units. Using the relationship
\be
\frac{\pi}{2g^3 G_N^{(5)}} = N^2
\ee
and also \cite{Peet:1998wn,Bianchi:2001de} that the field theory cutoff is rescaled from the one used here by a factor of $\sqrt{\lambda}$, that is $R/\delta_{\rm SG} \sim R/\delta\sqrt{\lambda}$, we arrive at 
\be\label{eq.GravityUniversalScaling}
I_n = - \frac{Q^2}{4} N^2 \ln \left( R/\delta \right)- \frac{Q^2}{8} N^2 \ln \lambda 
\ee
matching the field theory result (\ref{eq.FreeEnergyFieldTheory}) precisely.
Additionally, from (\ref{eq.EuclideanAction}) we get the scaling relation
\be\label{eq.EuclidActionScale}
I_n = \frac{(n+1)^2}{4n} I_1 \,.
\ee
However, it would be more correct to say that we worked in a particular renormalization scheme in which the entire action scales like (\ref{eq.EuclidActionScale}). More generally it is only the universal part, i.e. the coefficient of the logarithm, that obeys such an identity. The scaling results here translate into the results for the supersymmetric R\'enyi entropies in section (\ref{sec.SUSYRenyi}) upon using the definition (\ref{eq.singularPartitionFunction}).
\subsection{Holographic Wilson loop}\label{eq.HolographicWilsonLoop}
We wish to find the action associated with a Wilson loop in the fundamental representation holographically, using the relationship
\be
S_{\rm string}=-\ln W_n.
\ee 
To do this, we find the on-shell action of a stationary string solution
\be\label{eq.StringWorldSheet}
S_{\rm string}=\frac{1}{2\pi\alpha'}\int d^2s \,e^{-\Phi}\sqrt{\det G_{MN}\partial_\alpha X^M\partial_\beta X^N},
\ee
in the uplifted type IIB background given in section 2.2. As we shall see below, it is not sufficient to study the string embedding merely in the five dimensional effective theory, as this will lead to the wrong result for the Wilson loop. This illustrates a subtle point about the use of consistent truncations in holography: while computations concerning the supergravity sector, such as the value of the Euclidean partition function, are by definition consistently calculated by the truncated lower-dimensional theory, stringy probes can be sensitive to the full ten-dimensional background, and not just through the dependence of the action (\ref{eq.StringWorldSheet}) on the dilaton. In fact $\Phi$, the IIB dilaton, is constant in our background,
\be
e^{-\Phi}=1.
\ee
The non-zero $A^I$ in the metric components would induce an angular velocity for the angles on the $S^3$. The $S^3$ degenerates to a point at 
\be
\cos\xi=0,
\ee so the string should sit at this point on the internal manifold in order to preserve the $SU(2)$ $R$-symmetry. This corresponds to the string being uncharged under the $R$-symmetry. The string configuration is therefore given by a worldsheet spanning the $r$ and $t$ directions, which we will take to be the worldsheet coordinates:
\be
s_1=t,\, s_2=r,
\ee
such that
\be
\xi(r,t)=\frac{\pi}{2}.
\ee 
We can choose the angles on the $S^3$ and $\tau$,
\be
\tau(r,t)=\theta(r,t)=\phi(r,t)=\psi(r,t)=0, 
\ee
 without loss of generality. This loop, wrapping the $t$ direction, corresponds to the $\varphi$ loop of \cite{Hama:2012bg}, for which the large $N$ localization result was calculated in section 2.2.2. In terms of the induced worldsheet metric
\be
\gamma_{\alpha\beta}\equiv G_{MN}\partial_\alpha X^M\partial_\beta X^N,
\ee
the Nambu-Goto equations of motion in this curved background are given by
\be
\partial_\alpha\left(\sqrt{-\det \gamma}\,\gamma^{\alpha\beta}G_{MN}\partial_\beta X^N\right)+\frac{1}{2}\sqrt{-\det\gamma}\,\gamma^{\alpha\beta}\partial_\alpha X^N \partial_\beta X^P \partial_{X^M}G_{NP}=0.
\ee
With our ansatz and background, the equations of motion are indeed satisfied. The associated on-shell string world-sheet action is then given by
\begin{align}
S_{\rm string} &= \lim_{\Lambda\rightarrow\infty}\frac{1}{2\pi\alpha'}\int^{2 \pi n/g}_0 dt \int^{\Lambda}_{r_H} dr H^{-\frac{1}{3}}\Delta^{\frac{1}{2}}\nn
&=\lim_{\Lambda\rightarrow\infty}\frac{n}{g\alpha'}\left(\Lambda-r_h\right)\nn
&=-\frac{n+1}{2\alpha' g^2},
\end{align}
where we have subtracted the infinite counterterm from the UV end of the string. Crucially the measure gets a contribution $\Delta^{\frac{1}{2}} = H^{\frac{1}{3}}$ from the $\mS^5$ part of the geometry (\ref{eq.SU2Uplift}). Without considering the full ten-dimensional geometry of the string, one would thus not be able to obtain the correct Wilson-loop expectation value.

Using the relation $\sqrt{\lambda}=\frac{1}{g^2\alpha'}$ , this can be expressed in terms of field theory quantities as
\be
\ln W_n = \frac{n+1}{2}\sqrt{\lambda}.
\ee 
The corresponding scaling relations are given by
\be
\ln W_n = \frac{(n+1)}{2}\ln W_1,
\ee
and, for the Wilson loop contribution to the R\'enyi entropy,
\be
S^{(W)}_n=\frac{1}{2}\ln W_1.
\ee
The quantities agree precisely with those obtained by localization after taking the large $N$ limit. Again, the scaling is the same as that obtained in bulk dimension $d+1=4$ in \cite{Nishioka:2013haa}.

\section{Discussion}
In this paper we have found the exact expression for the supersymmetric R\'enyi entropy across a spherical entangling surface in ${\cal N}=4$ SYM theory to all orders in $N, \lambda$. We were able to obtain this result by mapping the computation of the R\'enyi entropy to that of a partition function on a four-dimensional ellipsoid. This calculation in turn can be performed using supersymmetric localization \cite{Hama:2012bg}. At large-$N, \lambda$ the field-theory result can be reproduced by a holographically dual calculation involving a supersymmetric black hole solution of Romans' ${\cal N}=4^+$ truncation of five-dimensional gauged supergravity. This precise match opens a number of interesting questions.

Firstly, the quantity we calculated in this work is a natural supersymmetric generalization of the usual information-theoretic R\'enyi entropy. In that case the R\'enyi entropies give one access to a fine-grained description of the entanglement properties of the state or ensemble of states. It would be interesting to understand whether the super-R\'enyis also contain useful information about the entanglement of the state.

Further, concerning the properties of the supersymmetric R\'enyi entropy, it is striking that the universal content of our four-dimensional computation has resulted in identical expressions as those previously obtained in three dimensions \cite{Nishioka:2013haa,Huang:2014gca,Nishioka:2014mwa}, once normalized by the entanglement entropy. For example, the universal part of the R\'enyi entropy obeys the scaling relation
\be
{\cal S}_n = \frac{3n+1}{4n}{\cal S}_1\,.
\ee
It would be interesting to understand whether supersymmetric R\'enyi entropies in dimensions greater than two take this form quite generally, similarly to the universal result of R\'enyi entropies in 2D CFT \cite{Calabrese:2004eu}.

Our computation implies a number of interesting relationships between seemingly unrelated field theories. Both on the field-theory side and on the gravity side our work shows a surprising duality with an a priori completely different object. Namely the partition function of the ${\cal N}=2^*$ massive deformation of ${\cal N}=4$ theory evaluated on $\mS^4$ \cite{Pestun:2007rz,Bobev:2013cja}. In fact the results of our paper, at large $N, \lambda$, can be mapped into those of the ${\cal N}=2^*$ theory by the simple replacement
\be
\frac{Q^2}{4} \longleftrightarrow 1 + M^2\,,
\ee
where $M$ is the mass parameter of the ${\cal N}=2^*$ theory.
This is particularly intriguing since we were able to obtain the gravity result by an analytic computation in ${\cal N}=4^+$ theory, while \cite{Bobev:2013cja} numerically solved the BPS equations of a newly constructed truncation of maximally gauged supergravity in five dimensions. We believe that the correspondence extends to all $N$ and $\lambda$  and it would be enlightening to understand the reason for the agreement between the two calculations, both on the field-theory side and for their bulk duals.

Finally, in a similar vein, one would like to understand the bulk representation of the deformation independence of the boundary field theory, discussed in section \ref{sec.branchedIsSquashed}, which ensures that the localized partition function on different geometries are equal to each other. In our work, we saw explicitly that the bulk partition function of a non-abelian black hole in non-abelian $SU(2)\times U(1)$ gauged supergravity precisely coincides with that of the maximally abelian $U(1)^3$ gauged supergravity in 5D \cite{Huang:2014pda}. This result is easy to understand at the classical level: we showed that the $SU(2)$ gauge field can be entirely oriented along the Cartan generator, and then explicitly embedded in the abelian theory (for an analogous story involving monopoles see \cite{Goddard:1976qe}). However, as soon as fluctuations are taken into account, {\it i.e.} moving away from the classical limit, one would expect the theories to differ. Yet, we know from the field theory analysis that the {\it full} quantum partition functions exactly coincide. The question is thus: how is this non-trivial equivalence encoded in the bulk?

\subsection*{Acknowledgements}
We would like to thank Hong Liu, Joe Minahan, Duff Neill, Eric Perlmutter and Silviu Pufu for helpful conversations and correspondence. JS acknowledges the hospitality the of LANL theory division. JS and ED acknowledge the hospitality of the Santa Fe Institute while this paper was being finalized. This work was supported in part by the U.S. Department of Energy (DOE) under cooperative research agreement Contract Number DE-FG02-05ER41360. 

\appendix
\section{Supersymmetry of the ${\cal N}=4^+$ theory}\label{app.SUSY}
After accounting for the signature change from mostly minus to mostly plus (remember $\gamma$ matrices have to be multiplied by $i$) and the necessary rescaling $g^2_{\rm Romans} = 8g^2$ to put the relevant sector of \cite{Romans:1985ps} into the form (\ref{eq.N=4plusLag}), we find the following fermion variations \cite{Romans:1985ps}
\bea\label{eq.FermiVariations}
\delta\chi_a &=& \frac{1}{\sqrt{2}}i\gamma^\mu\left(\partial_\mu\phi\right) \varepsilon_a+ A_{ab}\varepsilon^b + \frac{1}{2\sqrt{6}}\gamma^{\mu\nu}H_{\mu\nu ab}\varepsilon^b\,,\nn
\delta \psi_{\mu a} &=&D_\mu \varepsilon_a + i \gamma_\mu T_{ab}\varepsilon^b + \frac{i}{6\sqrt{2}}\left(\gamma_\mu{}^{\nu\rho} + 4 \delta_\mu{}^\nu \gamma^\rho  \right)H_{\nu\rho a b}\varepsilon^b\,,
\eea
where
\be
D_\mu \varepsilon_a = \nabla_\mu \varepsilon_a + \frac{g}{\sqrt{2}}A_\mu^I\left( \Gamma_{I45}\right)_{a}{}^b\varepsilon_b
\ee
and
\bea
X &=& e^{-\sqrt{2/3}\phi}\,,\nn
A_{ab} &=& \frac{g}{\sqrt{3}}\left(X - X^{-2}  \right)\left(\Gamma_{45}\right)_{ab}\,,\nn
T_{ab} &=&\frac{g}{3}\left( X + \tfrac{1}{2}X^{-2} \right) \left(  \Gamma_{45}\right)_{ab}\nn
H_{\mu\nu a b}&=&\frac{1}{2X}F_{\mu\nu}^I\left( \Gamma_I \right)_{ab}\,.
\eea
The matrices $\Gamma_i$ are $USp(4)\cong Spin(5)$ Euclidean gamma matrices
\be
\left(\Gamma_i \right)_a{}^b\left(\Gamma_j \right)_b{}^c + \left(\Gamma_j \right)_a{}^b\left(\Gamma_i \right)_b{}^c  = 2 \delta_{ij}\delta_a^c
\ee
with $\Gamma_I = \Gamma_i$ for $i=1,2,3$. We also have
\be
\left\{ \gamma^\mu\,,\,\gamma^\nu \right\} = 2 g^{\mu\nu}\,.
\ee
In this appendix, we adopt the convention to write position-space components of gamma matrices as $\gamma_t,\gamma_r\,\ldots$, while tangent-space components are numbered $\gamma_0,\gamma_1\,\ldots$. We also adopt and obvious tensor-product notation: matrices such as $\gamma_\mu (\Gamma_{3})_a{}^b$ act in an obvious way on the spin and $USp(4)$ indices of $\varepsilon^a$. This action can be concisely written as $\gamma_\mu\otimes \Gamma_3 \varepsilon$, now suppressing all spinorial indices.
\subsection{Killing Spinor}
We now want to solve for the bulk Killing spinor explicitly. For this purpose it is useful to record the non-vanishing components of the spin connection
\bea
\omega^{01} &=& \left(\frac{f'}{2H} - \frac{2fH'}{3H^2}\right)dt\,,\nn
\omega^{12}&=&-\sqrt{f} \left( 1 + \frac{rH'}{3H} \right)du\,,\nn
\omega^{13}&=&-\sqrt{f} \left( 1 + \frac{rH'}{3H} \right)\sinh u d\theta\,,\nn
\omega^{14}&=& -\sqrt{f} \left( 1 + \frac{rH'}{3H} \right)\sinh u \sin \theta d\varphi\,,\nn
\omega^{23}&=& - \cosh u d\theta\,,\nn
\omega^{34}&=& -\cos\theta d\varphi
\eea
We find that if we impose the condition (\ref{eq.FermiVariations}), as well as the additional algebraic constraint
\be\label{eq.Projector2}
\tilde P \varepsilon = 0 \,\qquad {\rm with}\qquad \tilde P =  \mathbf{1} - i \Gamma_{345} \,,
\ee
the components of the Killing spinor equation become
\bea
\left( \partial_t + \frac{g}{2} \left( 1 - \sqrt{2}i\mu \right)\right)\varepsilon &=&0\,,\nn
\left( \partial_r + \frac{ig}{3\sqrt{f}} \left(1 + \tfrac{H}{2}  \right)\gamma_1\otimes \Gamma_{45} + \frac{H'}{3H\sqrt{f}}\gamma_0 \otimes \Gamma_3  \right)\varepsilon &=&0\,,\nn
\left(\partial_u -\tfrac{1}{2}\gamma_0\gamma_1\gamma_2 \otimes \Gamma_3 \right)\varepsilon &=&0\,,\nn
\left(\partial_\theta -\tfrac{1}{2}\sinh u \gamma_0\gamma_1\gamma_3 \otimes \Gamma_3  - \tfrac{1}{2}\cosh u \gamma_2 \gamma_3\right)\varepsilon &=& 0\,,\nn
\left(\partial_\varphi -\tfrac{1}{2}\sinh u \sin\theta\gamma_0\gamma_1\gamma_4 \otimes \Gamma_3  - \tfrac{1}{2}\cosh u\sin\theta \gamma_2 \gamma_4 - \tfrac{1}{2}\cos\theta \gamma_3 \gamma_4\right)\varepsilon &=& 0.\nn
\eea
The solution to systems of equations of this type is given in general terms in \cite{Romans:1991nq}, whose notation we follow here. Matrices acting in the tensor product $\gamma\otimes\Gamma$ are abstractly denoted by $\overline{\Gamma}$. Since $\Gamma_{345}$ commutes with the condition (\ref{eq.Projector}), we can simply impose this projection independently, that is to say, we can work with a spinor $\varepsilon$ that satisfies (\ref{eq.Projector2}) from the outset and follow the steps in \cite{Romans:1991nq} without paying attention to the additional constraint. The temporal and angular equations, having no dependence on $r$, can be solved immediately, resulting in the expression
\bea
\varepsilon(r) &=& \exp\left(-\frac{gt}{2} \left( 1 - \sqrt{2}i\mu \right)  \right) \exp\left( \frac{u}{2}\gamma_{012}  \right)\nn
&& \times \exp\left( \frac{\theta}{2}\gamma_{23}   \right)\exp\left( \frac{\varphi}{2}\gamma_{34}  \right)\xi(r)\,.
\eea
Using the projection constraint (\ref{eq.Projector}) we can rewrite the radial equation in the form
\be
\partial_r \xi(r) = \left( a(r) + b(r) \overline{\Gamma}_1 \right)\xi(r)\,,
\ee
where we write the projector (\ref{eq.Projector}) as 
\be
P = \frac{1}{2} \left( 1 + x(r) \overline{\Gamma}_1 + y(r) \overline{\Gamma}_2 \right)\,,
\ee
where
\be
x(r) = -\frac{i}{\sqrt{f}}\,,\qquad  y(r) = \frac{grH}{\sqrt{f}}\,
\ee
and
\be
a(r) =\frac{1}{3 r H} \left(1 + \tfrac{1}{2} H \right)\,.
\ee
We have defined the matrices
\be
\overline{\Gamma}_1 = i \gamma_0 \otimes \Gamma_3\,,\qquad \overline{\Gamma}_2 = i \gamma_1 \otimes \Gamma_{45}
\ee
satisfying
\be
\overline{\Gamma}_1^2 = \overline{\Gamma}_2^2 = \mathbf{1}\,,\qquad \overline{\Gamma}_1\overline{\Gamma}_2 = -\overline{\Gamma}_2 \overline{\Gamma}_1\,.
\ee
The solution to this equation, given in\cite{Romans:1991nq}, is
\be\label{eq.KillSpinSol}
\xi(r) = \frac{1}{2}\left( u(r) + v(r) \overline{\Gamma}_2 \right) \left( 1 - \overline{\Gamma}_1 \right)(1 - i 1\otimes\Gamma_{345})\xi_0\,,
\ee
where we have also inserted the second, compatible, constraint on the spinor.
The functions work out to be
\be
u(r) = \sqrt{\frac{1+x(r)}{y(r)}}e^{w(r)}\,,\qquad v(r) = - \sqrt{\frac{1-x(r)}{y(r)}}e^{w(r)}\,,\qquad w(r) = \int^r a(r')dr'\,.
\ee
The two projections manifest in (\ref{eq.KillSpinSol}) make the black hole a $\tfrac{1}{4}$-BPS solution preserving four real supercharges.  Explicitly we find
\be
\int a(r') dr' = \frac{1}{6}\ln \left( r^3 H \right)\,,
\ee
so that the Killing spinor becomes
\be
\xi(r) = \frac{H^{-1/3}}{2 \sqrt{g}}\left(\sqrt{\sqrt{f} -i} \, - \,\overline {\Gamma}_1 \sqrt{\sqrt{f} + i}  \right)  \left(1 - \overline{\Gamma}_1 \right) \left( 1 - i \Gamma_{345} \right)\xi_0
\ee
in terms of an arbitrary constant symplectic Majorana spinor $\xi_0$.
\section{Embedding into $U(1)^3$ truncation}\label{sec.STUembedding}\label{sec.STU}
The recent preprint \cite{Huang:2014pda} finds the gravity dual of the field theory on the singular space in terms of a solution of five dimensional ${\cal N}=2$ $U(1)^3$ gauged supergravity, specifically the STU model.
In this appendix we describe the $U(1)^3$ truncation of the ${\cal N}=8$ $SO(6)$ gauged supergravity in five dimensions and comment on the relation to our work. This theory can be seen as arising from an $S^5$ reduction of type IIB supergravity in ten dimensions \cite{Cvetic:1999xp}. The action of the bosonic sector of the resulting ${\cal N}=2$ supergravity can be written as
\be\label{eq.STUaction}
S = \int \sqrt{-g}\left[ R + g^2 {\cal V} - \frac{1}{4}G_{ij}F_{\mu\nu}^{(i)}F^{\mu\nu (i)} - \frac{1}{2}G_{ij}\partial_\mu X^{(i)} \partial^\mu X^{(j)} \right] + S_{\rm CS}\,.
\ee
As in the case of the ${\cal N}=4^+$ theory we do not need the Chern-Simons part of the action as its contribution vanishes throughout the analysis of this paper. We work in units setting $16\pi G_{N}^{(5)}=1$. In addition to the metric $g$ and three vector fields $A^{(i)}$, there are three scalar fields $X^i$, satisfying the constraint
\be\label{eq.STUconstraint}
 X^1 X^2 X^3 = 1\,.
\ee
The moduli-space metric $G_{ij}$ is diagonal
\be
G_{ij} = \left(\begin{array}{ccc}(X^{(1)})^{-2} & 0 & 0 \\0 &(X^{(2)})^{-2}   & 0 \\0 & 0 & (X^{(3)})^{-2}\end{array}\right)
\ee
and the potential is
\be
{\cal V} =2 \sum_{i=1}^3(X^{(i)})^{-1}\,,
\ee
both evaluated on solutions of the constraint (\ref{eq.STUconstraint}), so that the theory only contains two independent scalar fields. We now discuss black hole solutions of this theory, which carry charges under all three $U(1)$ gauge fields \cite{Behrndt:1998jd}. 
\subsection{Hyperbolic three-charge Black Hole}
Here we review the salient features of the three-charge black hole solutions of theory (\ref{eq.STUaction}). These were found by  \cite{Behrndt:1998jd}, their BPS limit was discussed in \cite{Behrndt:1998ns} and their embedding into type IIB supergravity established in \cite{Cvetic:1999xp}.

The metric of the hyperbolic three-charge black hole can be written as
\be
ds^2 = -\frac{f(r)}{\left(H_1 H_2 H_3\right)^{2/3}}dt^2 + \left( H_1 H_2 H_3 \right)^{1/3} \left(\frac{dr^2}{f(r)} + r^2 ds^2(\mH^3)\right)
\ee
with
\be
f(r) = -1 - \frac{m}{r^2} + g^2 r^2 H_1 H_2 H_3\,,\qquad H_i = 1 + \frac{q_i}{r^2}\,.
\ee
The gauge fields $A^i$ as well as scalars $X^i$ take the form
\be
A^i_r =\mu_i -\frac{i\tilde q_i}{r^2 + q_i},,\qquad X^i = H_i^{-1}\left(H_1 H_2 H_3  \right)^{1/3}
\ee
One often sees the notation $X^1=S\,,X^2 = T\,,X^3=U$ and the theory being referred to as the STU model. The solution of interest to us is the two equal charge solution, which is a special case of the above, with $\tilde q_i = q_i$, and $q_1 = q_2 = q\,,q_3=0$. The BPS solution, preserving half of the super symmetries further has $m=0$. We denote $H_1 = H_2 :=H$. It is manifest that the two equal charge solution of this theory can be identified with the special solution (\ref{eq.STUBlackHole}) of $SU(2)\times U(1)$ gauged supergravity considered in this paper. In order to understand this embedding better, is instructive to study the the full ten-dimensional geometry, as a solution to the low-energy limit of type IIB string theory. This has been worked out before \cite{Cvetic:1999xp}, so we will present only the necessary ingredients here.  We refer the reader to \cite{Cvetic:1999xp} for more details. 

\subsubsection{Uplift to Type IIB}
The five dimensional theory (\ref{eq.STUaction}) corresponds to a consistent truncation of type IIB on $S^5$ with non-trivial five form flux and constant dilaton. The only relevant ingredient for the present analysis is the uplift formula of the metric
\be\label{eq.10Duplift}
ds_{10}^2 = \sqrt{\Delta} ds_5^2 + \frac{1}{g^2 \sqrt{\Delta}}\sum_{i=1}^3 X_i^{-1}\left( d\mu_i^2 + \mu_i^2 \left( d\phi_i^2 + g A^i \right)^2 \right)\,,
\ee
with
\be
\mu_1 = \sin\theta\,,\qquad \mu_2 = \cos\theta \sin\psi\,,\qquad \mu_3 = \cos\theta \cos\psi\,,
\ee
and
\be
\Delta = \sum_{i=1}^3 X_i \mu_i^2\,.
\ee
By setting $A^3=0$ and $A^1 = A^2$, one can show that the five-sphere uplift becomes identical to that of $SU(2)\times U(1)$ theory, Eq. (\ref{eq.SU2Uplift}), with $A^I=0$ for $I=1,2$, i.e. a gauge field entirely in the Cartan direction of $SU(2)$. This explains why the two bulk duals give the same answer. As remarked upon in the discussion, it is less obvious that the two theories should agree away from the classical limit.

Since this uplift coincides with the uplift of the $SU(2)\times U(1)$ gauged supergravity, when the gauge field is in the Cartan, one sees immediately that the Wilson loop expectation value is also matched by a string embedded in (\ref{eq.10Duplift}).
\bibliography{Renyi}{}
\bibliographystyle{utphys}

\end{document}